\documentclass[prl,twocolumn,superscriptaddress]{revtex4}%
\usepackage{amsfonts}
\usepackage{amsmath}
\usepackage{amssymb}
\usepackage{graphicx}%
\usepackage[dvipsnames]{xcolor}
\newcommand{\minus}{\scalebox{0.75}[1.0]{$-$}}

\begin{document}

\title{Revealing the complex nature of bonding in binary high-pressure compound FeO$_2$}

\author{E. Koemets} 
\affiliation{Bayerisches Geoinstitut, University of Bayreuth, D-95440 Bayreuth, Germany}
\affiliation{Institut Charles Gerhardt Montpellier (UMR CNRS 5253), Universit\'e de Montpellier, F-34095, Montpellier cedex 5, France}

\author{I. Leonov}
\affiliation{Institute of Metal Physics, Sofia Kovalevskaya Street 18, 620219 Yekaterinburg GSP-170, Russia}
\affiliation{Materials Modeling and Development Laboratory, NUST ``MISIS'', 119049 Moscow, Russia}
\affiliation{Ural Federal University, 620002 Yekaterinburg, Russia}

\author{M. Bykov}, 
\affiliation{Bayerisches Geoinstitut, University of Bayreuth, D-95440 Bayreuth, Germany}

\author{E. Bykova}
\affiliation{Bayerisches Geoinstitut, University of Bayreuth, D-95440 Bayreuth, Germany}
\affiliation{Carnegie Institution of Washington, Earth and Planets Laboratory, 5241 Broad Branch Road, N.W., Washington, DC, 20015, USA}

\author{S. Chariton}
\affiliation{Bayerisches Geoinstitut, University of Bayreuth, D-95440 Bayreuth, Germany}

\author{G. Aprilis}
\affiliation{Material Physics and Technology at Extreme Conditions, Laboratory of Crystallography, Universit\"at Bayreuth, D-95440 Bayreuth, Germany}
\affiliation{The European Synchrotron Radiation Facility, 38043 Grenoble Cedex 9, France}

\author{T. Fedotenko}
\affiliation{Material Physics and Technology at Extreme Conditions, Laboratory of Crystallography, Universit\"at Bayreuth, D-95440 Bayreuth, Germany}

\author{S. Cl\'ement}
\affiliation{Laboratoire Charles Coulomb (L2C) - UMR CNRS 5221 Universit\'e de Montpellier. CC069, 34095 Montpellier, France}

\author{J. Rouquette}
\affiliation{Institut Charles Gerhardt Montpellier (UMR CNRS 5253), Universit\'e de Montpellier, F-34095, Montpellier cedex 5, France}

\author{J. Haines}
\affiliation{Institut Charles Gerhardt Montpellier (UMR CNRS 5253), Universit\'e de Montpellier, F-34095, Montpellier cedex 5, France}

\author{V. Cerantola}
\affiliation{The European Synchrotron Radiation Facility, 38043 Grenoble Cedex 9, France}

\author{K. Glazyrin}
\affiliation{Photon Science, Deutsches Elektronen-Synchrotron, D-22607 Hamburg, Germany}

\author{C. McCammon}
\affiliation{Bayerisches Geoinstitut, University of Bayreuth, D-95440 Bayreuth, Germany}

\author{V. B. Prakapenka}
\affiliation{Center for Advanced Radiation Sources, University of Chicago, Chicago, Illinois 60437, USA}

\author{M. Hanfland}
\affiliation{The European Synchrotron Radiation Facility, 38043 Grenoble Cedex 9, France}
 
\author{H.-P. Liermann}
\affiliation{Photon Science, Deutsches Elektronen-Synchrotron, D-22607 Hamburg, Germany}

\author{V. Svitlyk}
\affiliation{The European Synchrotron Radiation Facility, 38043 Grenoble Cedex 9, France}

\author{R. Torchio}
\affiliation{The European Synchrotron Radiation Facility, 38043 Grenoble Cedex 9, France}

\author{A. D. Rosa}
\affiliation{The European Synchrotron Radiation Facility, 38043 Grenoble Cedex 9, France}
 
\author{T. Irifune}
\affiliation{Geodynamics Research Center, Ehime University, 2-5 Bunkyo-cho, Matsuyama 790-8577, Japan}
 
\author{A. V. Ponomareva}
\affiliation{Materials Modeling and Development Laboratory, NUST ``MISIS'', 119049 Moscow, Russia}

\author{I. A. Abrikosov}
\affiliation{Department of Physics, Chemistry and Biology (IFM), Link\"oping University, SE-581 83 Link\"oping, Sweden}

\author{N. Dubrovinskaia} 
\affiliation{Material Physics and Technology at Extreme Conditions, Laboratory of Crystallography, Universit\"at Bayreuth, D-95440 Bayreuth, Germany}
\affiliation{Department of Physics, Chemistry and Biology (IFM), Link\"oping University, SE-581 83 Link\"oping, Sweden}

\author{L. Dubrovinsky}
\affiliation{Bayerisches Geoinstitut, University of Bayreuth, D-95440 Bayreuth, Germany}

\begin{abstract}

Extreme pressures and temperatures are known to drastically affect the chemistry of iron oxides resulting in numerous compounds forming homologous series $n$FeO$\cdot m$Fe$_2$O$_3$ and the appearance of FeO$_2$. Here, based on the results of \emph{in situ} single-crystal X-ray diffraction, M\"ossbauer spectroscopy, X-ray absorption spectroscopy, and DFT+dynamical mean-field theory calculations we demonstrate that iron in high pressure cubic FeO$_2$ and isostructural FeO$_2$H$_{0.5}$ is ferric (Fe$^{3+}$), and oxygen has a formal valence less than two. Reduction of oxygen valence from 2, common for oxides, down to 1.5 can be explained by a formation of a localized hole at oxygen sites.

\end{abstract}
\maketitle

At ambient (or low) pressures, three different iron oxides are known: Fe$_2$O$_3$ with a mineral name hematite; Fe$_3$O$_4$ magnetite -- the oldest known magnetic material, and FeO w\"ustite, which is non-stoichiometric and typically iron deficient. At extreme pressures and temperatures, the synthesis yields numerous iron oxides with unexpected compositions (such as Fe$_4$O$_5$, Fe$_5$O$_6$, Fe$_7$O$_9$, Fe$_5$O$_7$, Fe$_{25}$O$_{32}$, etc.), unusual crystal structures, and intriguing physical properties, demonstrating the complexity of the binary Fe-O system \cite{Ref_1,Ref_2,Ref_3,Ref_4,Ref_5}. 
It was suggested that iron oxides at high-pressure and high-temperature conditions (HP-HT) could be systematized by homologous structural series $n$FeO$\cdot m$Fe$_2$O$_3$ formed by oxygen (O$^{2-}$) and iron in ferrous and/or ferric states (Fe$^{2+}$ and Fe$^{3+}$, correspondingly). Besides the end-members, iron could exist in the mixed-valence state in this series (intermediate between 2+ and 3+ valence formally), defined by the stoichiometry of HP iron oxides. However, the recent finding of cubic FeO$_2$ (space group $Pa\overline{3}$), and closely related FeO$_2$H$_x$ ($x$ up to 1) phases \cite{Ref_6,Ref_7,Ref_8} suggests that not only iron but also oxygen could have a variable oxidation state in iron oxides (or oxyhydroxides). 

Powder X-ray diffraction (PXRD) \cite{Ref_6}, X-ray Absorption Spectroscopy (XAS, \cite{Ref_9,Ref_10}), and Nuclear Forward Scattering (NFS) studies \cite{Ref_10} of cubic high-pressure FeO$_2$H$_x$ ($x=0$ to 1) compounds as well as results of some theoretical works \cite{Ref_11,Ref_12} were used to argue that iron is ferrous in these phases even at strongly oxidized conditions and thus oxygen forms peroxide (O$_2$)$^{2-}$ ions. However, the question concerning the oxidation state of both iron and oxygen in FeO$_2$ and FeO$_2$H$_x$ remains controversial, primarily because of harsh experimental conditions and ambiguous results. For example, while XAS data were interpreted to indicate that iron is ferrous \cite{Ref_9,Ref_10}, yet NFS data of cubic FeO$_2$ \cite{Ref_10} shows center shifts ($\sim$0.15 mm/s at 80 GPa) that are unrealistic for any ferrous compound. The HP-HT synthesis of the FeO$_2$H$_x$ phase for the XAS analysis in the work \cite{Ref_9} was performed from FeOOH using KCl as a pressure transmitting medium, which could chemically react \cite{Ref_13} and the reaction products could affect interpretation of spectra collected for FeO$_2$H$_x$ phase. Available experimental information on the crystal structure of FeO$_2$ and FeO$_2$H$_x$ phases is based on PXRD \cite{Ref_6,Ref_8}, which makes the analysis of the Fe-O and O-O distances unreliable compared to more complex yet more informative structural refinements from single-crystal X-ray diffraction data (SC-XRD) \cite{Ref_14}. Additionally, some theoretical works \cite{Ref_15,Ref_15a,Ref_16,Ref_17} suggest that iron is ferric in FeO$_2$ illustrating the necessity of performing high-accuracy experiments to establish the physical and chemical properties of this phase. 

\begin{figure}[t]
\includegraphics[width=0.5\textwidth]{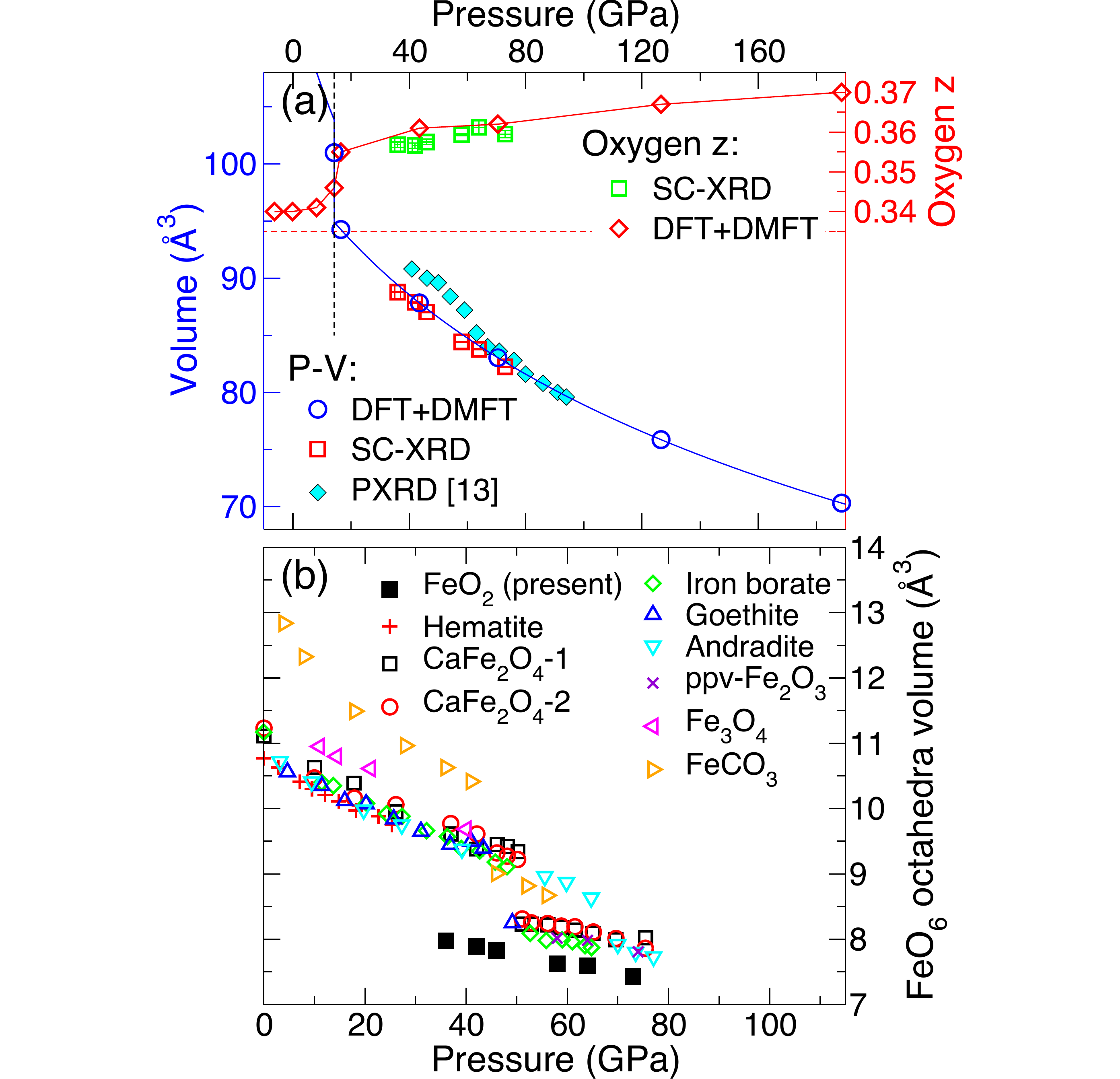}
\caption{(Color online) Compressional behavior of HP-PdF$_2$-type FeO$_2$: (a) lattice volume and fractional $z$ coordinate of oxygen under pressure as obtained by DFT+DMFT at $T=1160$ K in comparison to XRD data. Note that below $\sim$14 GPa, FeO$_2$ shows a low-spin to high-spin phase transition, associated with the formation of local moments. (b) Pressure dependence of volumes of the FeO$_6$ octahedra in various compounds according to Ref.~\cite{Ref_30} and references therein.}
\label{Fig_1}
\end{figure}

The goals of the present work are to clarify the high-pressure crystal chemistry of cubic FeO$_2$ and FeO$_2$H$_x$ phases and determine the oxidation state of iron and oxygen. These are not only of importance as a fundamental problem for HP-HT chemistry but also highly relevant for geosciences \cite{Ref_6}. In order to achieve the goals of our studies, we perform multi-method synchrotron-based experiments including advanced \emph{in situ} SC-XRD, X-ray absorption and M\"ossbauer spectroscopy using laser-heated diamond anvil cells (see Supplementary Material \cite{Ref_Suppl}, Table~S1). We support our experimental results by the DFT+dynamical mean-field theory (DFT+DMFT) calculations \cite{Ref_19,Ref_20} of the electronic structure, magnetic and valence state of iron. In addition, we perform a full structural optimization and compute the crystal structure parameters of paramagnetic FeO$_2$ under pressure within DFT+DMFT \cite{Ref_21,Ref_22,Ref_22a,Ref_23,Ref_24,Ref_25,Ref_26}. 
Our experimental and theoretical results suggest that iron in HP-PdF$_2$-type FeO$_2$ and FeO$_2$H$_x$ is ferric $(3+)$. We show the absence of a molecular (O$_2$)$^{3-}$ bonding state in HP-PdF$_2$-type FeO$_2$, implying that the oxidation state of oxygen is equal to $1.5\minus$ due to oxygen-metal negative charge transfer. Such a charge transfer is expected to shorten the Fe-O distance and consequently reduce the volume of FeO$_6$ octahedra, which should cause both iron polyhedra and the entire structure to become highly incompressible. We propose that in major phases constituting the lower mantle and the core-mantle boundary the oxidation state of oxygen may significantly deviate from $2\minus$ due to this effect.


Compression of iron in oxygen at ambient temperature to 25(1) GPa did not produce any chemical reaction, but laser heating of the sample at this pressure to $\sim$1500(100) K led to the formation of Fe$_2$O$_3$ (space group $R\minus3c$, lattice parameters $a = 6.271(7)$ \AA, $c = 7.662(4)$ \AA) (see Fig.~S2), in agreement with literature data \cite{Ref_27}. After further compression to 46(2) GPa, the laser heating of a sample was performed at $\sim$1200(100) K. The XRD pattern of the temperature-quenched product drastically changed. The XRD analysis shows cubic FeO$_2$ with the space group $Pa\overline{3}$ and unit cell parameter $a = 4.4313(14)$ \AA, which is close to the values previously reported for ``pyrite-type'' FeO$_2$ \cite{Ref_6,Ref_16}. Iterative heating of the samples at different pressures resulted in the growth of micro-crystals of cubic FeO$_2$ that enabled performing an \emph{in situ} SC-XRD data collection, the structure solution, and further refinements with the HP-PdF$_2$-type structure in a range of pressures from 36(1) to 73(2) GPa, (see Table~S1\cite{Ref_Suppl}). The results on the crystal structure refinements from \emph{in situ} SC-XRD datasets of the FeO$_2$, which have never been reported before, are summarized in Table~S2\cite{Ref_Suppl}; the illustration of corresponding crystal structure of HP-PdF$_2$ FeO$_2$ is presented in Fig.~S1\cite{Ref_Suppl}; the compressional behavior of this phase is presented in Fig.~\ref{Fig_1}. Structural analysis suggests that the shortest O-O bond length (in the O-O dimer) varies from 2.213(7) to 2.104(15) \AA\ within the pressure range from $\sim$36 to 73 GPa (see Table~S2). For peroxides (in molecular or crystalline forms), the distances between the closest oxygen atoms at ambient pressure are very characteristic -- from about 1.2 to 1.5 \AA\ \cite{Ref_28} (e.g., in MgO$_2$ it is about 1.492 \AA\ at ambient pressure) and under compression these distances should not increase. In the case of FeO$_2$, such a large observed value for the shortest O-O distance suggests, from crystal-chemical point of view, the absence of chemical bonding between these oxygen atoms. Even in the case of the structural model of FeO$_2$ refined against powder XRD data (``pyrite-type FeO$_2$'' model \cite{Ref_6}), the shortest O-O bond is $\sim$1.896 \AA\ at 76 GPa which is too large for peroxides. 

\begin{figure}[t]
\includegraphics[width=0.5\textwidth]{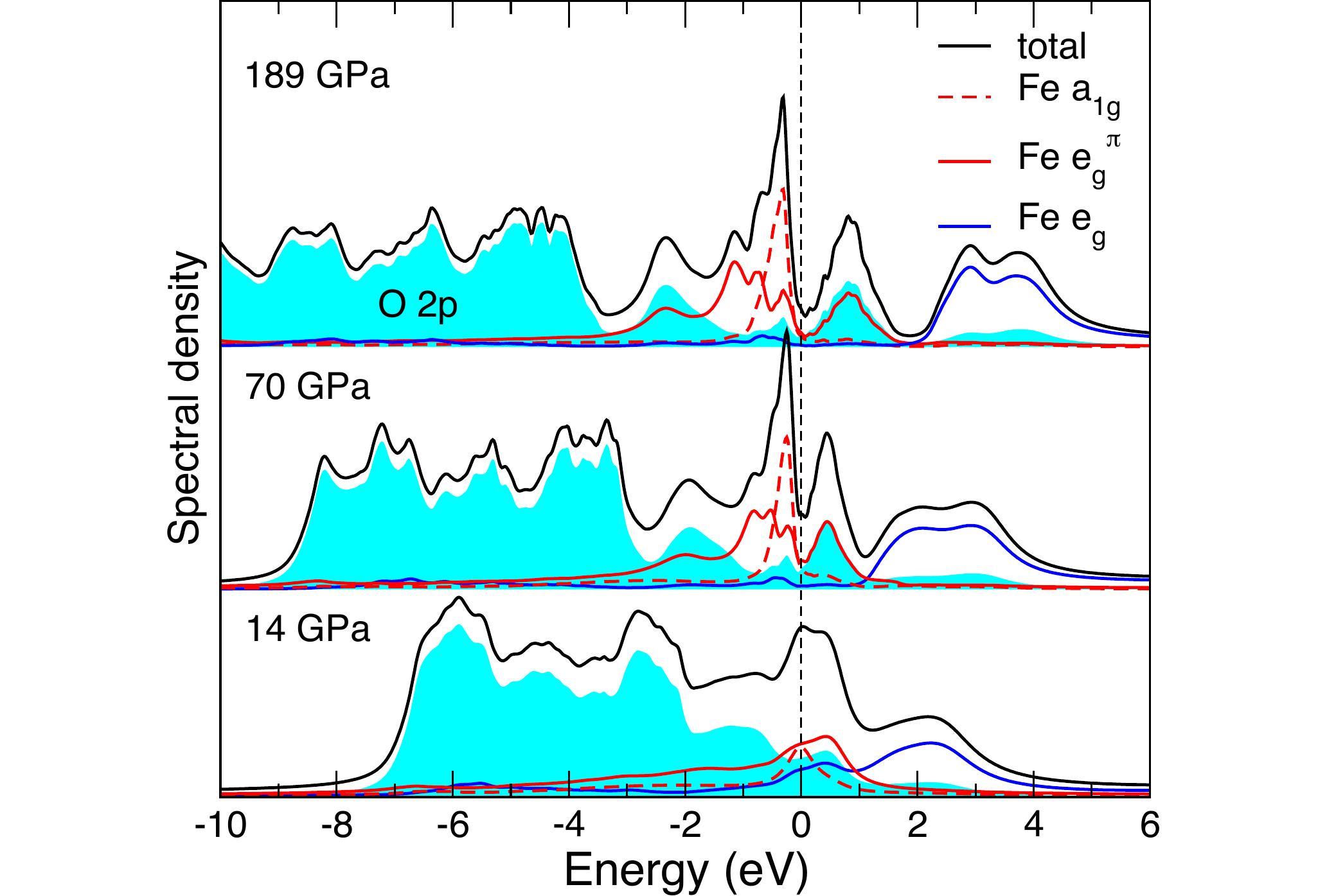}
\caption{(Color online) Orbitally-resolved spectral functions of HP-PdF$_2$-type FeO$_2$ for the Fe $3d$ ($a_{1g}$, $e_g^\pi$ and $e_g^\sigma$) and O $2p$ (blue shaded area) states obtained by DFT+DMFT at 1160 K as a function of compression.}
\label{Fig_2}
\end{figure}

A number of transition metals, for example, osmium and ruthenium, the neighbors of iron in the VIIIb group of the Periodic Table, form dioxides, OsO$_2$ and RuO$_2$ with the HP-PdF$_2$-type structure \cite{Ref_29,Ref_30} (space group $Pa\overline{3}$). The shortest O-O distance in these dioxides is equal to $\sim$2.5 \AA\ at ambient conditions. These phases are characterized by low compressibility (for details see Refs.~\cite{Ref_29,Ref_30}), and cubic FeO$_2$ is very incompressible as well, according to our experimental data (see Fig.~\ref{Fig_1}(b)). Thus, accounting for the relatively long O-O distances in FeO$_2$ at HP, one could expect that it adopts rather the HP-PdF$_2$-type structure than forms a peroxide. Additionally, according to the ``rule of a thumb'' \cite{Ref_31}, the behavior of compounds (particularly oxides) of an element at high pressure is similar to that of compounds of the elements with higher atomic number in the same group of the Periodic Table at low pressures. 

The results of structural studies and crystal-chemical considerations are consistent and pointing towards highly unusual crystal chemistry of Fe-O bonds in HP-PdF$_2$-type FeO$_2$. In fact, Streltsov \emph{et al.} \cite{Ref_15} have suggested on the basis of \emph{ab initio} calculations for FeO$_2$ that the valence of iron is $3+$ and have classified the material as lying ``in between'' oxides and peroxides with the anion described as (O$_2$)$^{3-}$. However, the structural model obtained from the powder diffraction analysis reported by Hu \emph{et al.} \cite{Ref_6} have been used in the calculations, and the input crystal structure of cubic FeO$_2$ has not been optimized self-consistently \cite{Ref_15}. While other electronic structure studies used the DFT+U method to compute the crystal structure parameters of FeO$_2$ \cite{Ref_11,Ref_12,Ref_17}, these computations assume the existence of a long-range magnetic order in HP-PdF$_2$ phase of FeO$_2$, in contradiction with experiment. 
As a result, such computations cannot give a reliable results for the shortest O-O bond length in FeO$_2$, predicting either a highly unusual increase of the O-O distance under pressure \cite{Ref_11} or a large O-O distance of 2.232 \AA\ at 76 GPa \cite{Ref_17}.

We resolve this point by computing the crystal structure phase stability and electronic structure of the paramagnetic HP-PdF$_2$ phase of FeO$_2$ using a fully charge self-consistent DFT+DMFT method (see Suppl. Material~\cite{Ref_Suppl}). Within DFT+DMFT we perform a full structural optimization of the lattice parameters of the \emph{paramagnetic} HP-PdF$_2$ phase of FeO$_2$ and compare our results with experimental data obtained through more precise \emph{in situ} SC-XRD. In Fig.~\ref{Fig_1} we display our results for the crystal structure parameters obtained by DFT+DMFT. In contrast to the previous DFT+U results \cite{Ref_11}, we observe that upon compression the O-O distance decreases from 2.286 \AA\ at 17 GPa to 2.085 \AA\ at 70 GPa, which is in close agreement with our SC-XRD data. Indeed, our SC-XRD measurements give $\sim$2.213(7) \AA\ at 36(1) GPa and 2.117(8) \AA\ at 73(2) GPa. Our DFT+DMFT calculations show that at a pressure of $\sim$70 GPa FeO$_2$ is a poor metal (see Fig.~\ref{Fig_2}) with about 5.21 electrons in the Wannier Fe $3d$ states (4.07 electrons inside the atomic sphere with radius $\sim$0.78 \AA, in accord with a bond valence analysis). Oxygen states are partially occupied with $\sim$0.61 hole states in the Wannier O $2p$ orbitals. The local magnetic moment is $\sim$1.59 $\mu_\mathrm{B}$ (fluctuating local moment of 0.83 $\mu_\mathrm{B}$). Our results for the decomposition of electronic state into atomic configurations (valence states) show that the valence value for Fe is nearly $3+$ at $\sim$70 GPa: Fe$^{3+}$ $3d^5$ configuration has a weight of about 50\%, with a $\sim$30\% admixture of the Fe$^{2+}$ $3d^6$ state ($\sqrt{0.5}|d^5\rangle+\sqrt{0.3}|d^6\rangle$, see Fig.~S7). 

\begin{figure}[t]
\includegraphics[width=0.5\textwidth]{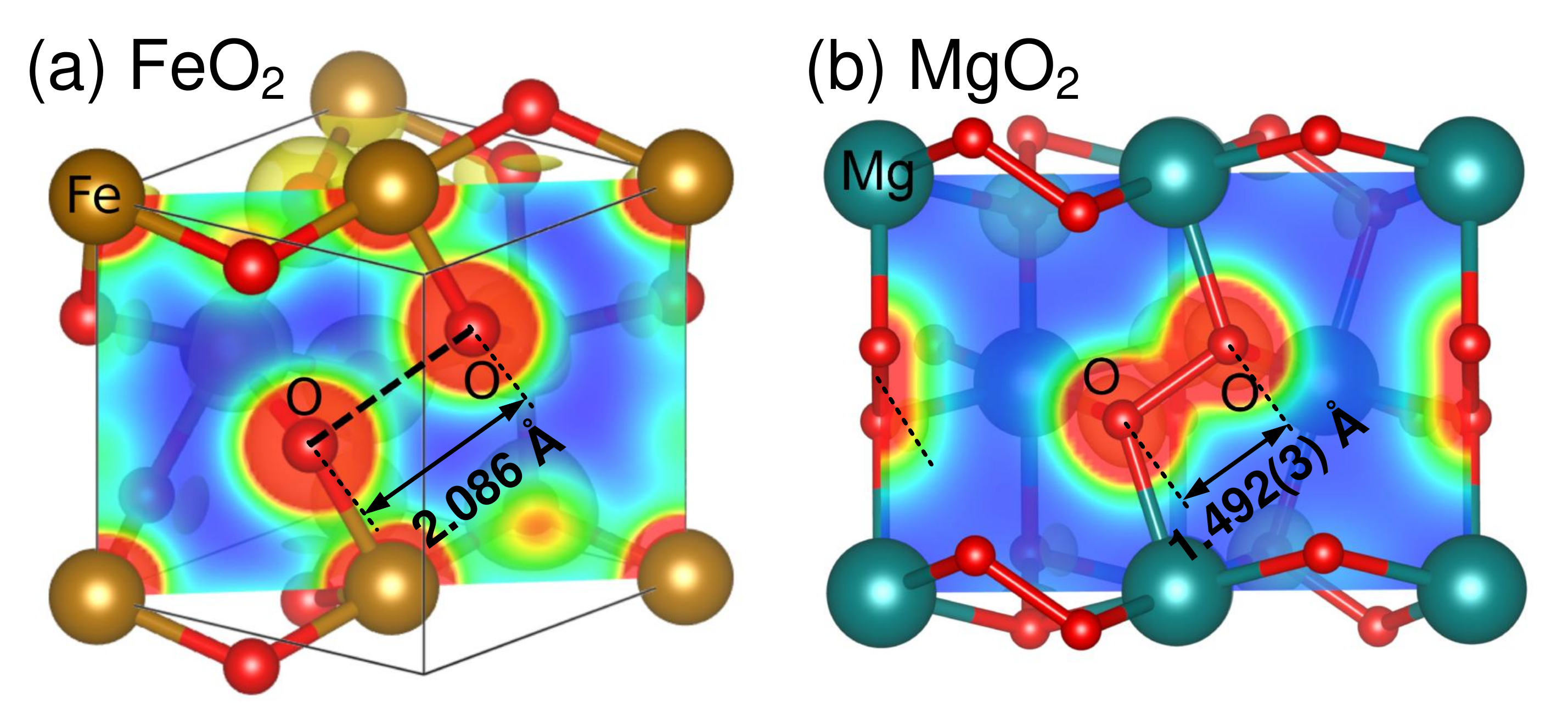}
\caption{(Color online) (a) Crystal structure and valence electron density plot for HP-PdF$_2$-type FeO$_2$ obtained by DFT+DMFT at $T=1160$ K and 70 GPa (unit cell volume 83.0 \AA$^3$, the calculated O-O distance 2.086 \AA\ is too large for the peroxide-type structure). (b) DFT (HSE03) results for pyrite-type MgO$_2$ peroxide at ambient conditions (uni cell volume 113.3 \AA$^3$ and the O-O distance is about 1.492(3) \AA\ \cite{Ref_40}).}
\label{Fig_3}
\end{figure}

In Fig.~\ref{Fig_2} we see that due to distorted FeO$_6$ octahedron symmetry, the Fe $t_{2g}$ states split into a $a_{1g}$ singlet and $e_g^\pi$ doublet. Fe $e_g^\sigma$ orbitals are empty and are located well above the Fermi level at $1\minus4$ eV. Fe $t_{2g}$ states form weakly renormalized $(m/m^*\sim1.6)$ quasiparticle bands near $E_\mathrm{F}$. 
No evidence for a metal-to-insulator phase transition in FeO$_2$ (below $\sim$189 GPa) was observed within our fully relaxed DFT+DMFT calculations \cite{Ref_11,Ref_12}, in agreement with experiment. In fact, under experimental setup presented here samples were black with a metallic shine, implying a metallic state of FeO$_2$. Moreover, within DFT+DMFT the low-to-high spin state transition is found to occur below $\sim$14 GPa, i.e., below the stability field of the HP-PdF$_2$-type FeO$_2$. We therefore propose that decomposition of FeO$_2$ experimentally observed below $\sim$30 GPa is associated with a change of the electronic state of iron in FeO$_2$. Most notably, our DFT+DMFT results confirm that even at $\sim$189 GPa the shortest O-O bond remains sufficiently large (1.86 \AA), implying the absence of covalent ``molecular'' O-O bonding in FeO$_2$.

The DFT+DMFT results agree well with our M\"ossbauer and X-ray absorption spectroscopy data that shows a low-spin state of nearly $3+$ iron ions in the studied pressure range (see below). Most importantly, our fully relaxed and charge self-consistent DFT+DMFT calculations lead to a different bonding picture of FeO$_2$ in comparison to the analysis by Streltsov \emph{et al.} \cite{Ref_15}. Our results reveal the absence of a molecular (O$_2$)$^{3-}$ bonding state, i.e., in FeO$_2$ iron has effective charge $3+$, and oxygen $1.5\minus$. We see that at about 70 GPa the bonding O-O $\sigma$ states appear at about $\minus2$ eV, while the anti-bonding $\sigma^*$ states due to mixing with the Fe $t_{2g}$ states at the Fermi level split into the $t_{2g} \pm \sigma^*$ combinations (seen as two peaks at $\minus0.5$ and $+0.5$ eV). Importantly, the empty $t_{2g}-\sigma^*$ O-O band is located $\sim$0.5 eV above the Fermi level, confirming the formation of a localized hole at the O sites. We conclude that FeO$_2$ belongs to the class of negative charge-transfer materials in the Zaanen-Sawatzky-Allen scheme (materials in which excitation energy for the transfer of electrons from the O $2p$ to Fe $3d$ states is negative) \cite{Ref_34,Ref_34a}. In such materials, instead of having an electronic configuration corresponding to the formal valence state, e.g. Fe$^{4+}$ ($3d^4$) and O$^{2-}$ ($2p^6$) configuration in FeO$_2$, the system prefers to have a configuration with higher occupation of the $3d$-shell, creating holes on oxygen.

At the same time, the bonding-antibonding splitting of the O $2p$ orbitals is small, just $\sim$2$\minus3$ eV, indicating negligible bonding between the two oxygen atoms. This agrees well with our analysis of the charge density distribution in FeO$_2$ in comparison to magnesium peroxide MgO$_2$ (space group $Pa\overline{3}$). Our results are summarized in Fig.~\ref{Fig_3}, highlighting the conclusion regarding the absence of a molecular (O$_2$)$^{3-}$ or (O$_2$)$^{2-}$ bonding state in FeO$_2$ in the studied interval of pressures. In fact, while MgO$_2$ clearly shows the formation of a ``molecular'' (O$_2$)$^{2-}$ bond with $\sim$21\% of a maximal electron density value in the center of the O-O bond, for cubic FeO$_2$ the value is only 5\% at 70 GPa (it increases to $\sim$8\% at 189 GPa, see Fig.~S7). Thus, in sharp contrast to MgO$_2$ no covalent O-O bond is seen for FeO$_2$ in Fig.~\ref{Fig_3}. Therefore, the results of explicit examination of the calculated electronic structure and charge density distribution in HP-PdF$_2$-type FeO$_2$ confirms our above conclusion on the absence of chemical bonding between these oxygen atoms, as well as that Fe oxidation state is nearly $3+$. 

Our M\"ossbauer spectroscopy data (see Fig.~S4) are consistent with iron in the Fe$^{3+}$ state \cite{Ref_3,Ref_35}. We note that the center shift that we obtained for cubic FeO$_2$ at 58(2) GPa (0.06(5) mm/s) is in good agreement with the value of Liu \emph{et al.} \cite{Ref_10}. Our experimental and theoretical results thus imply that the oxidation state of oxygen in HP-PdF$_2$-type FeO$_2$ is equal to $1.5\minus$ due to oxygen-metal negative charge transfer. Such a charge transfer is expected to shorten the Fe-O distance and consequently reduce the volume of FeO$_6$ octahedra, which should cause both iron polyhedra and the entire structure to become highly incompressible. Indeed, fitting the experimental SC-XRD pressure-volume data for cubic FeO$_2$ with the 3rd order Birch-Murnaghan equation of state (EoS) gave the EoS parameters with a large bulk modulus: $V_0=97.6(3)$ \AA$^3$, the unit cell volume; $K_0=305(9)$ GPa, the bulk modulus; and $K'=4.0$ (fixed), the pressure derivative of the bulk modulus. The compressibility of FeO$_6$ octahedra is low ($K_{0,\mathrm{octa.}}=350(4)$ GPa) and the octahedral volume is significantly smaller than that known for any other compound, including those with ferric iron in the low-spin state (Fig.~\ref{Fig_1}(b)).

\begin{figure}[t]
\includegraphics[width=0.45\textwidth]{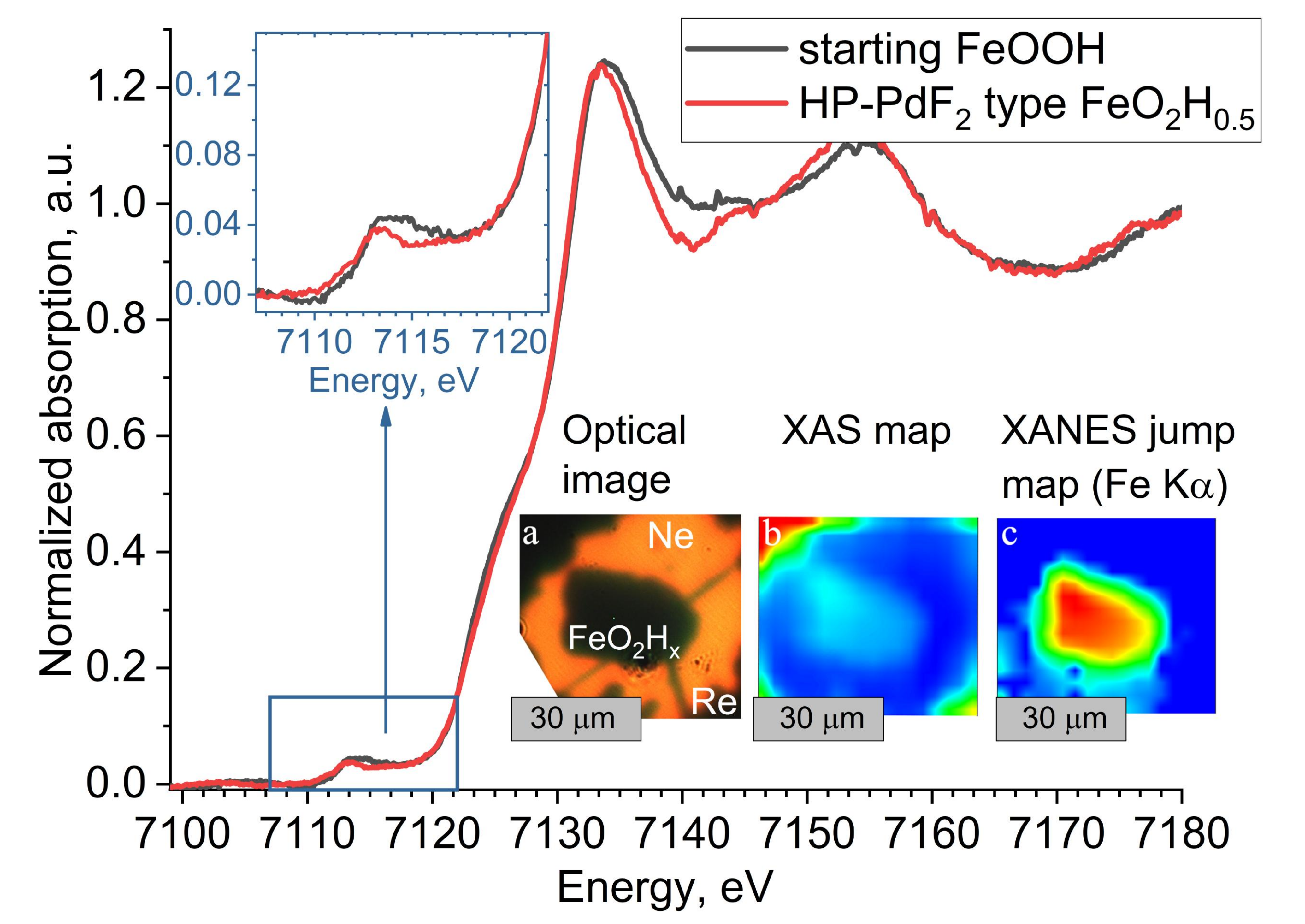}
\caption{(Color online) Normalized X-ray absorption spectra of Fe K$\alpha$ edge for FeO$_2$H$_{0.5}$ at 86(2) GPa (red line) synthesized from starting FeOOH (gray line) after laser-heating to 1700(200)K at at 86(2) GPa (DAC11, see Table~S1). 
Bottom right inset: (a) Microphotograph of synthesized FeO$_2$H$_{0.5}$ sample; (b) and (c) XAS absorption contrast and absorption jump maps of sample chamber (data collected with steps of 3 $\mu$m covering the whole sample chamber, palette reflects the relative values of the absorption and absorption jump).}
\label{Fig_4}
\end{figure}

We complimented studies of cubic FeO$_2$ by investigations of high-pressure behavior of iron (III) oxyhydroxide, FeO$_2$H$_x$. Laser heating of goethite, $\alpha$-FeOOH, at $\sim$81(2) GPa and 1500(100) K resulted in formation of a cubic phase with the lattice parameter $a=4.430(1)$ \AA. The structure was solved and refined from SC-XRD data (Table~S2\cite{Ref_Suppl}), and the arrangement of Fe and O was found to correspond to the HP-PdF$_2$-type structure, confirming that FeO$_2$ and FeO$_2$H$_x$ are isostructural phases. The lattice parameter suggests the chemical composition FeO$_2$H$_{0.4}$ \cite{Ref_36}. The relatively high value of the shortest O-O distance ($\sim$2.262(5) \AA) rules out the peroxide-type chemical bonding between oxygen atoms and the presence of hydrogen does not shorten this bond length. 

\begin{figure}[t]
\includegraphics[width=0.45\textwidth]{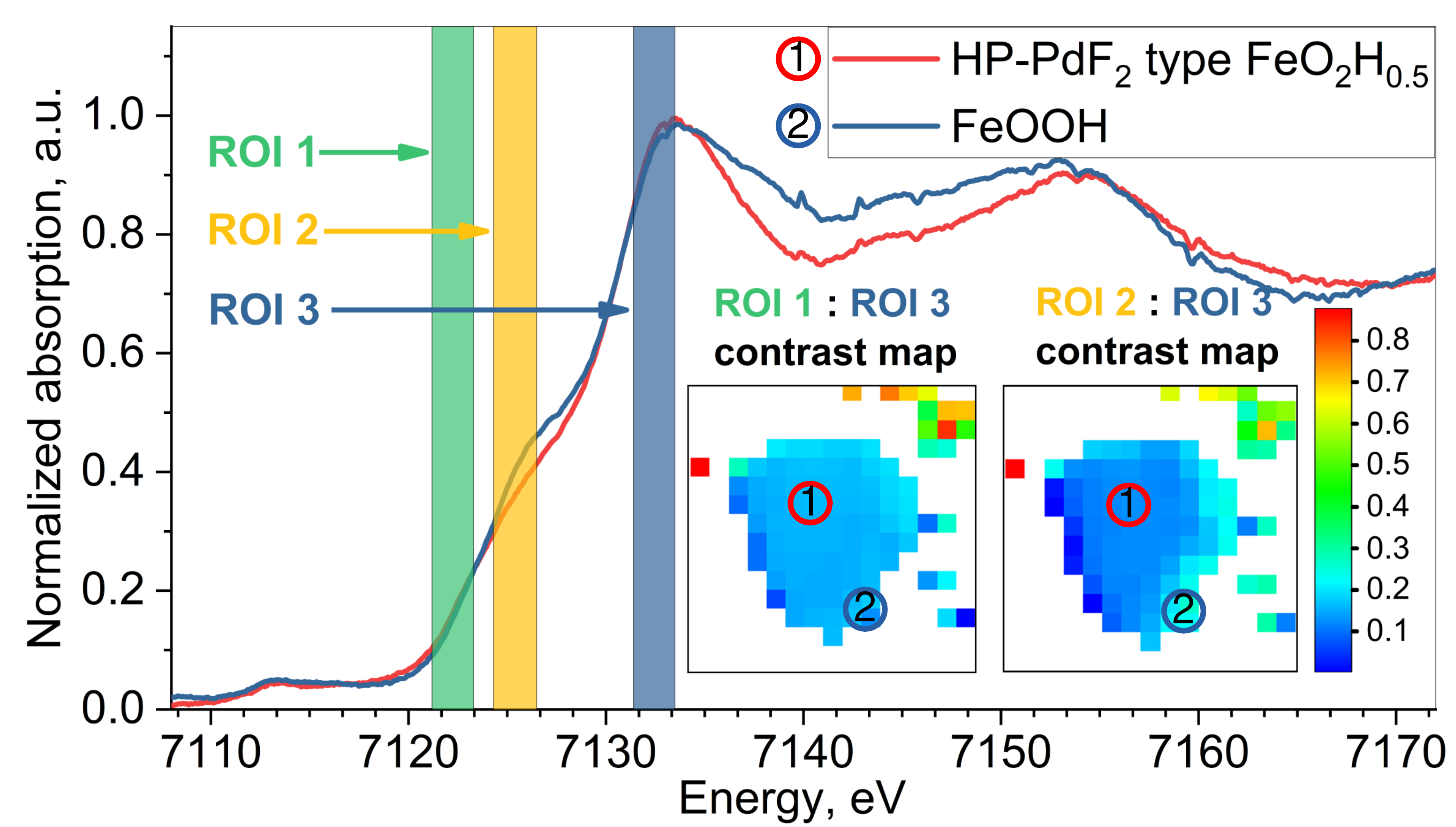}
\caption{(Color online) Fe K$\alpha$ edge XAS spectra and contrast maps of starting FeOOH and FeO$_2$H$_{0.5}$ synthesized by laser heating to 1700(200) K at 86(2) GPa (DAC11, see Table~S1). XAS spectra collected at different spots of the sample are denoted by red and blue circles with numbers (red line and point 1 corresponds to the cubic FeO$_2$H$_{0.5}$ and blue line and point 2 corresponds to the remaining FeOOH). The inset shows the contrast maps built for different ROIs (each pixel of the map corresponds to the spectra collected with 3 $\mu$m steps over the sample chamber).  The ranges of ROIs used to build contrast maps described in \cite{Ref_41} are highlighted in colors.}
\label{Fig_5}
\end{figure}

In order to confirm the oxidation state of iron in cubic FeO$_2$H$_x$, we performed \emph{in situ} XANES measurements on the sample synthesized by laser heating of goethite at 86(2) GPa and 1700(200) K in a DAC equipped with anvils made of polycrystalline diamonds \cite{Ref_37}. Powder XRD data confirmed the synthesis of material with the lattice parameter $a=4.449(5)$ \AA\ (Fig.~S5), which corresponds to the composition FeO$_2$H$_{0.5}$. In the XANES spectra collected in the center of a sample at the Fe K$_\alpha$ edge, the pre-edge peak narrows after synthesis of FeO$_2$H$_{0.5}$, and a negligible changes in the edge feature are observed; however, the position of the absorption jump remains the same for the starting FeOOH and cubic FeO$_2$H$_{0.5}$ (Fig.~\ref{Fig_4}), inferring that iron does not alter its oxidation state during this transformation and remains 3+. We have performed the XANES mapping of a whole sample with 3 $\mu$m step to verify the sample homogeneity. The comparative contrast maps \cite{Ref_41} were built for different energy Regions of Interest (ROI) ($\sim$7121 eV [ROI 1], $\sim$7124 eV [ROI 2] and $\sim$7131 eV [ROI 3] corresponding to the beginning of the edge, edge feature and the normalized first XANES peak correspondingly, see Fig.~\ref{Fig_5}). The contrast built for the [ROI 2]:[ROI 3] suggest small deviations of a XANES edge feature caused by the presence of negligible amount of FeOOH on the edges of a sample (the spectrum collected at point 2 in Fig.~\ref{Fig_5}) due to the temperature gradients during laser heating. In the same time, [ROI 1]:[ROI 3] conrast map, sensitive to the position of the edge, rules out the presence of Fe$^{2+}$ in the sample. Comparison of XANES data obtained from the starting FeOOH and FeO$_2$H$_{0.5}$ and the contrast maps unambiguously confirms that iron is ferric in both species (Figs.~\ref{Fig_4} and \ref{Fig_5}). Therefore, oxygen in FeO$_2$H$_{0.5}$ has the formal oxidation state $1.75\minus$.

Generalizing our observations on cubic HP-PdF$_2$-structured FeO$_2$ and FeO$_2$H$_x$ phases and taking into account that compounds with $x$ up to 1 have been described in the literature, we conclude that at pressures above $\sim$50 GPa, the oxidation state of oxygen can significantly deviate from $2\minus$. Experimental and theoretical results on cubic FeO$_2$ and FeO$_2$H$_x$ phases may be concise in terms of the concept of valence. For our purposes we accept a definition of the ``valence'' of an element as a measure of its combining power with other atoms when it forms chemical compounds, or, as in the bond valence model \cite{Ref_38}, as the number of electrons the atom uses for bonding. Thus, in the HP-PdF$_2$-type structured FeO$_2$ and FeO$_2$H$_{0.5}$ iron has the valence $3+$, and oxygen -- $1.5$ and $1.75$, correspondingly. Reducing the oxygen valence from 2, common for oxides, down to 1.5 can be explained by a formation of a localized hole at oxygen sites, which leads to a reduction of the Fe-O distance and, as a consequence, of the volume of FeO$_6$ octahedra.

In conclusions, our results show that iron (III) dioxide FeO$_2$ can form at $\sim$45 GPa, i.e., at significantly milder conditions than reported earlier ($\sim$75 GPa) \cite{Ref_6}, which corresponds to the pressures at the depth of $\sim$1150 km in the upper part of the Earth's lower mantle. However, the high oxidation state of iron in HP-PdF$_2$-type cubic FeO$_2$ and FeO$_2$H$_x$ phases makes their appearance hardly possible for any plausible equilibrium homogeneous lower mantle assemblages with the oxygen fugacity below iron-w\"ustite buffer \cite{Ref_39}.

\begin{acknowledgments}
The authors acknowledge the Deutsches Elektronen-Synchrotron (DESY, PETRA III), the European Synchrotron Radiation Facility (ESRF), and the Advance Photon Source (APS) for provision of beamtime. N.D. and L.D. thank the Federal Ministry of Education and Research, Germany (BMBF, grant no. 05K19WC1) and the Deutsche Forschungsgemeinschaft (DFG projects DU 954–11/1, DU 393–9/2, and DU 393–13/1) for financial support. N.D. thanks the Swedish Government Strategic Research Area in Materials Science on Functional Materials at Link\"oping University (Faculty Grant SFO-Mat-LiU No. 2009 00971).
Electronic structure calculations were supported by the Russian Science Foundation (Project No. 18-12-00492). Theoretical analysis of chemical bonding was supported by the Ministry of Science and Higher Education of the Russian Federation in the framework of Increase Competitiveness Program of NUST ``MISIS'' (No. K2-2019-001) implemented by a governmental decree dated 16 March 2013, No. 211. 
Support from the Knut and Alice Wallenberg Foundation (Wallenberg Scholar Grant No. KAW-2018.0194), the Swedish Government Strategic Research Areas in Materials Science on Functional Materials at Link\"oping University (Faculty Grant SFO-Mat-LiU No. 2009 00971) and SeRC, and the Swedish Research Council (VR) grant No. 2019-05600 is gratefully acknowledged.
\end{acknowledgments}

\end{document}